\documentclass[conference]{IEEEtran}
\IEEEoverridecommandlockouts
\usepackage{cite}
\usepackage{amsmath,amssymb,amsfonts}
\usepackage{algorithmic}
\usepackage{graphicx}
\usepackage{textcomp}
\usepackage{xcolor}
\def\BibTeX{{\rm B\kern-.05em{\sc i\kern-.025em b}\kern-.08em
    T\kern-.1667em\lower.7ex\hbox{E}\kern-.125emX}}

\usepackage{authblk}
\usepackage{algorithm}
\usepackage{multirow} 

\usepackage[colorlinks=true, linkcolor=blue, citecolor=blue, urlcolor=blue]{hyperref}

\def\mbf{\mathbf}
\def\beq{\begin{equation}}
\def \eeq{\end{equation}}
\def\bbmat{\begin{bmatrix}}
\def\ebmat{\end{bmatrix}}

\def\R{{\mathbb{R}}}

\def\Ncu{N_{cu}}
\def\Nap{N_{ap}}
\def\Ntg{N_{tg}}
\def\setA{\mathcal{A}}
\def\setAt{\mathcal{A}_t}
\def\setAr{\mathcal{A}_r}
\def\setU{\mathcal{U}}
\def\setT{\mathcal{T}}

\def\Nrfa{N_{\mathrm{RF}}^{(a)}}
\def\setBin{\{0, 1\}}
\def\up{u^{\prime}}
\def\rhoauupbar{\bar{\rho}_{u, u^\prime}^a}
\def\gaubar{\bar{g}_{a,u}}
\allowdisplaybreaks

\begin{document}

\title{ASSENT: Learning-Based Association Optimization for Distributed Cell-Free ISAC
\thanks{This work was supported by the National Science Foundation (NSF) under Grant CCF-2322191.}
}

\author{Mehdi Zafari}
\author{A. Lee Swindlehurst}
\affil{Department of Electrical Engineering and Computer Science, University of California, Irvine}

\maketitle



\begin{abstract}

Integrated sensing and communication (ISAC) is a key emerging 6G technology. 
Despite progress, ISAC still lacks scalable methods for joint access point (AP) clustering and user/target scheduling in distributed deployments under fronthaul limits. 
Moreover, existing ISAC solutions largely rely on centralized processing and full channel state information, limiting scalability.
This paper addresses joint AP clustering, user and target scheduling, and AP mode selection 
in distributed cell-free ISAC systems operating with constrained fronthaul capacity.
We formulate the problem as a mixed-integer linear program (MILP) that jointly captures interference coupling, RF chain limits, and sensing requirements, providing optimal but computationally demanding solutions.
To enable real-time and scalable operation, we propose ASSENT (ASSociation and ENTity selection), a graph neural network (GNN) framework trained on MILP solutions to efficiently learn association and mode-selection policies directly from lightweight link statistics.
Simulations show that ASSENT achieves near-optimal utility while accurately learning the underlying associations. 
Additionally, its single forward pass inference reduces decision latency compared to optimization-based methods.
An open-source Python/PyTorch implementation with full datasets is provided to facilitate reproducible and extensible research in cell-free ISAC. 

\end{abstract}



\section{Introduction}
\label{sec:intro}

Integrated sensing and communication (ISAC) is emerging as a core paradigm for next-generation wireless systems, simultaneously serving  communication users (CUs) and environmental sensing needs using shared spectrum and hardware resources~\cite{liu2022dual-functional}.
Cell-free architectures~\cite{ngo2017cellfree}, which eliminate predefined cell boundaries and enable cooperation among access points (APs), were originally proposed for communication networks to enhance spectral efficiency and manage interference more effectively.
This architecture is now being extended to ISAC systems~\cite{umut2025cellfreeisac}, where it offers additional advantages for sensing, such as larger effective apertures, improved clutter mitigation, and higher parameter estimation accuracy.

Depending on the level of coordination and computation, cell-free architectures can either be implemented with a central server (CS) that has access to global information and handles all decision-making, or in a distributed manner, where computation is decentralized among APs and only limited information is exchanged over the network.
Although centralized cell-free ISAC architectures have demonstrated strong performance in small-scale or controlled scenarios, their reliance on full channel state information (CSI), unlimited fronthaul capacity, and global coordination makes them unsuitable for large-scale, real-world deployments.

Distributed cell-free ISAC architectures provide a scalable alternative by leveraging spatially distributed APs equipped with local computation capabilities.
In such systems, APs operate using their local information, such as CSI and sensed target parameters, and exchange only limited information with a CS acting as a controller and aggregator.
In large-scale deployments, allowing all APs to jointly serve all CUs and targets is computationally prohibitive.
Hence, the network must cluster subsets of APs for cooperation, schedule subsets of CUs and targets to be served, and determine the operation mode of each AP, whether it transmits, receives, or (in principle) operates in full-duplex mode.
Therefore, realizing a distributed cell-free ISAC network introduces a set of new design challenges, including distributed resource allocation, AP clustering, user and target scheduling, and mode selection, all under practical fronthaul and hardware constraints.

Several recent studies have leveraged distributed optimization techniques to address resource allocation in cell-free networks~\cite{zafari2024admm, zafari2025decentralized}.
However, a key bottleneck in extending these architectures to ISAC lies in the absence of efficient association and entity selection mechanisms that can jointly determine which APs serve which CUs and targets, and in which operational modes.
The scheduling problem for ISAC systems has been investigated in a few recent works assuming full network knowledge~\cite{time-domain-scheduling-2024, setareh2025scheduling}.
In parallel, graph learning-based methods have emerged as promising tools for modeling cooperative ISAC networks~\cite{peng2025graph, wang2024GNN}.
Nevertheless, existing approaches typically assume centralized computation with global information, treat communication and sensing tasks in isolation, overlook spatial interference among CUs, or often rely on heuristic strategies that lack adaptability to dynamic and large-scale network deployments.

In this work, we consider a distributed cell-free ISAC network with constrained fronthaul capacity.
To limit the fronthaul load, each AP reports low-dimensional statistics for its communication and sensing links to a CS.
These lightweight statistics inform an association problem that consists of joint AP clustering, user and target scheduling, and AP mode selection, which is formulated as a mixed-integer linear program (MILP).
To capture network dynamics and enable real-time decision-making, we propose \textit{\textbf{ASS}ociation and \textbf{ENT}ity selection} (ASSENT), a graph neural network (GNN)-based framework trained on offline solutions of the MILP that learns the desired association behavior.
Once trained, ASSENT performs fast forward inference to deliver real-time decisions, shifting the heavy optimization burden offline and avoiding per-slot MILP implementation.
We also provide an open-source Python/PyTorch implementation of the proposed framework, together with the training and evaluation datasets. 
The contributions of this work are summarized below:
\begin{itemize}
\item \textbf{Association Optimization}: We formulate a joint AP clustering, user/target scheduling, and AP mode selection problem as an MILP for cell-free ISAC systems operating under constrained fronthaul capacity.
\item \textbf{ASSENT}: We propose ASSENT, a GNN-based learning framework that learns the optimal association decisions supervised by the optimal MILP solutions.
\item \textbf{Open-Source Implementation}: We develop and release a publicly available Python and PyTorch-based implementation of the MILP formulation and the ASSENT framework to support reproducible and extensible research.
\item \textbf{Evaluation}: We evaluate the performance of the MILP and the ASSENT framework, and analyze the effectiveness of multiple GNN architectures (NNConv, GATv2, TransformerConv) for association learning.
\end{itemize}



\section{System and Channel Model}
\label{sec:sys-mod}

Let us consider a cell-free network consisting of $\Nap$ distributed APs indexed by set $\setA$, $\Ncu$ CUs indexed by set $\setU$, and $\Ntg$ targets indexed by set $\setT$.
Each AP $a$ is equipped with an antenna array of size $M_a$, with a limited number of RF chains $\Nrfa$, where $\Nrfa \leq M_a$.
The system employs orthogonal frequency-division multiplexing (OFDM) and operates in time-division duplex (TDD) mode.
A resource block (RB) is defined as a contiguous portion of time-frequency resources consisting of $Q_f$ subcarriers in the frequency domain and $Q_t$ OFDM symbols in the time domain.
We consider a downlink scenario in which a subset $\setAt$ of APs transmits data to CUs and illuminates targets for sensing, while another subset $\setAr$ of APs operates in receive mode to capture the reflected target echo signals.
We consider half-duplex operation mode for the APs, i.e., $\setAt \cap \setAr = \emptyset$, to avoid self-interference.

\subsection{Communication Channel Model}

To model the communication channels, we adopt a narrowband stochastic channel based on the 3GPP Urban Micro (UMi) street canyon model~\cite{3gpp_r18}.
For each user-AP link, a distance-dependent line-of-sight (LoS) probability is defined according to the UMi scenario, and the LoS state determines the corresponding path-loss. The statistical link parameters such as the mean and standard deviation (STD) of the angle spreads, shadow fading, and the Rician K-factor are drawn following the specifications in~\cite{3gpp_r18}.
A correlated shadowing model is employed to capture spatial coupling among CUs.
The covariance between the shadow fading variables of CUs $j$ and $k$ as seen from AP $a$ is expressed as $\text{cov}(\psi_{a,j}, \psi_{a,k}) = \sigma_{a,j} \sigma_{a,k} \exp(-d_{j,k}/D_\text{corr})$, where $\psi_{a,j}$ denotes the shadow fading random variable, $\sigma_{a,j}$ its STD, $d_{j,k}$ the inter-user distance, and $D_{\text{corr}}=50\,\mathrm{m}$ the decorrelation distance.

For the non-LoS (NLoS) component, the spatial correlation matrix $\mathbf{R}_{a,u}$ is modeled assuming a uniform circular array.
The power angular spectrum is characterized by a truncated Laplacian distribution in both azimuth and elevation domains: $P(\phi|\mu_\phi, \sigma_\phi) = C \exp (-\sqrt{2} |\phi - \mu_\phi|_\text{wrap} / \sigma_\phi)$, where $\mu_\phi$ and $\sigma_\phi$ denote the mean and STD of the angular spread, respectively.
Using these stochastic parameters, the channel vector between user $u$ and AP $a$ on symbol $p$ and subcarrier $q$ is given by $\mathbf{h}_{a,u}[p,q]$.
To emulate imperfect channel estimation, the estimated channel at AP $a$ is modeled as $\widehat{\mbf{h}}_{a,u}[p,q] = \mbf{h}_{a,u}[p,q] + \mbf{n}[p,q]$, where $\mathbf{n}[p,q]$ represents the additive error.


The association decisions are determined by the CS on a per-RB basis using the lightweight statistics and limited information reported by the APs.
The instantaneous channel gain between AP $a$ and user $u$ on resource element $(p,q)$ is denoted by $g_{a,u}[p,q] = \|\widehat{\mbf{h}}_{a,u}[p,q]\|^2$.
The average channel gain over a given RB $r$ is then obtained as
\begin{equation}
    \label{eq:channel-gain-RB}
    \bar{g}_{a,u} = \frac{1}{Q_t} \frac{1}{Q_f} \sum_{(p,q)\in r} g_{a,u}[p,q],
\end{equation}
which serves as a compact channel statistic to be shared with the CS instead of the full CSI.
Note that since the analysis is per RB, the RB index $r$ is omitted for notational simplicity.
In addition to channel gains, each AP $a$ also computes the spatial correlation between the channels of CU $u$ and $\up$, defined as
\begin{equation}
    \rho_{u, \up}^a [p,q] = \frac{|\widehat{\mbf{h}}^H_{a,u}[p,q] \, \widehat{\mbf{h}}_{a,\up}[p,q]|^2}{\|\widehat{\mbf{h}}_{a,u}[p,q]\|^2 \cdot \|\widehat{\mbf{h}}_{a,\up}[p,q]\|^2},
\end{equation}
and its average over the given RB is measured analogously to~\eqref{eq:channel-gain-RB} and denoted by $\bar{\rho}_{u, \up}^{a}$.
Consequently, each AP $a$ reports to the CS only a low-dimensional set of real-valued parameters for communication link characterization: the vector of average channel gains $\mbf{g}_a = \{\bar{g}_{a,u}\}_{u\in \setU} \in \R^{\Ncu}$, and the matrix of spatial user correlations $\mbf{S}_a = \{\bar{\rho}_{u, \up}^{a}\}_{u,\up \in \setU} \in \R^{\Ncu \times \Ncu}$, to be used for the association optimization.

\subsection{Sensing Channel Model}

For the sensing links, we assume the same UMi street canyon environment as for CUs. Target $t$ is located at distance $d_{a,t}$ from AP $a$. The LoS probability between $a$ and $t$ is determined using the 3GPP distance-dependent LoS model.
To characterize the target radar cross section (RCS), we adopt the statistical measurements reported in~\cite{nextGalliance_tr_phase2}.
Each AP $a$ reports to the CS the following parameters for each target $t$: 3D AP-target distance $d_{a,t}$, estimated RCS $\bar{\beta}_{a,t}^{\text{rcs}}$ averaged over a given RB, and estimated LoS state, inferred from either the distance or prior measurements.
At the CS, for each bistatic sensing link formed by a transmit AP $a_t$, a target $t$, and a receive AP $a_r$, an approximate channel gain is computed as
\begin{align}
    \bar{g}_{a_t, t, a_r} = \ &\frac{G_\text{tx} G_\text{rx} \, \lambda_w^2 \, {\bar{\beta}}_{a_t,t,a_r}^\text{rcs}}{(4\pi)^3 \, d_{a_t, t}^2 \, d_{t,a_r}^2} \cdot W(p_{a_t,t}^\text{LoS}) W(p_{t,a_r}^\text{LoS}),
\end{align}
where $G_\text{tx}$ and $G_\text{rx}$ denote the antenna gains of the transmitting and receiving APs, respectively, $\lambda_w$ is the carrier wavelength, and ${\bar{\beta}}_{a_t,t,a_r}^\text{rcs} = (\bar{\beta}_{a_t,t}^\text{rcs} + \bar{\beta}_{t, a_r}^\text{rcs}) / 2$ represents the effective RCS obtained by averaging the transmit-target and target-receive links.
The gains and the wavelength are known at the CS, and $\bar{\beta}_{a,t}^\text{rcs}$, $d_{a,t}$, and $p_{a,t}^\text{LoS}$ are provided by APs for all links.
The term $W(p^\text{LoS})$ is a weighting function that modulates the contribution of each link based on its LoS probability:
\begin{equation}
    W(p^\text{LoS}) = w_\text{NLoS} + (w_\text{LoS} - w_\text{NLoS}) p^\text{LoS},
\end{equation}
where $w_\text{LoS}$ and $w_\text{NLoS}$ are tunable parameters controlling the relative influence of LoS and NLoS links.
The resulting $\bar{g}_{a_t,t,a_r}$ serves as a bistatic channel quality indicator and is subsequently used in the association optimization process.



\section{Association Optimization Problem}
\label{sec:problem}

In large-scale distributed cell-free ISAC networks, clustering and scheduling are fundamental tasks required for scalability and efficiency.
Prior work often decouples these steps, first forming clusters of APs and then scheduling CUs and targets within those clusters.
However, this overlooks the strong coupling between the AP-user/target associations and overall network interference patterns.
In particular, clustering decisions that are agnostic to user/target scheduling may underutilize network resources or create conflicting interference zones.
This problem becomes more pronounced in distributed architectures where each AP has only local CSI and coordination with the CS is limited due to fronthaul constraints.
Therefore, we propose a joint clustering and scheduling formulation that determines, in a single step, (i) which APs serve which CUs, (ii) which APs illuminate which targets, and (iii) which APs receive echoes from scheduled targets, using lightweight statistics at the CS such as average channel gains and inter-user spatial correlation.
This joint design enables more efficient use of network resources and ensures compatibility with distributed implementations that cannot rely on full CSI or global scheduling coordination.

\subsection{Optimization Variables}

The association between AP $a$ and CU $u$ is represented by a binary variable $x_{a,u} \in \setBin$, where $x_{a,u} = 1$ indicates that AP $a$ serves CU $u$ in the downlink, and $x_{a,u} = 0$ otherwise.
For target $t$, a binary association variable $y_{a,t}^\text{tx} \in \setBin$ is defined to indicate whether AP $a$ is transmitting to illuminate target $t$ ($y_{a,t}^\text{tx} = 1$) or not, and a variable $y_{a,t}^\text{rx} \in \setBin$ is defined to indicate whether AP $a$ is receiving echo signals from target $t$ ($y_{a,t}^\text{rx} = 1$) or not.
The operation mode of AP $a$ is also determined by variable $\tau_a \in \setBin$, where $\tau_a = 1$ indicates transmit mode and $\tau_a = 0$ receive mode.
Additionally, a binary variable $s_t \in \setBin$ is introduced to determine whether target $t$ is scheduled ($s_t = 1$) or not.
If scheduled, target $t$ must be associated with two non-overlapping subsets of APs, $\setAt$ for illumination and $\setAr$ for echo reception.

\subsection{Objective Function}

The objective of the association problem is to maximize a weighted sum of utility functions for CUs and targets.
For each $k \in \mathcal{U} \cup \mathcal{T}$, representing either a CU or a target, a weight $\lambda_k$ is assigned based on its service requirements and historical scheduling frequency, ensuring fairness and preventing any entity from being persistently under-served. 
The communication utility function is accordingly defined as
\begin{align}
    U_\text{comm} &= \sum_{u\in\setU} \lambda_u \sum_{a\in\setA} (\gaubar - \nu \sum_{\up \neq u} \rhoauupbar \gaubar x_{a,\up}) x_{a,u}  \label{eq:ucomm}\\
     & \triangleq \sum_{u\in\setU} \sum_{a\in\setA} \lambda_u (\gaubar x_{a,u} - \nu \sum_{\up \neq u} \rhoauupbar \gaubar v_{a,u,\up}), \nonumber 
\end{align}
in which associations likely to introduce interference to other scheduled CUs are penalized, and the hyperparameter $\nu \geq 0$ tunes the severity of the penalty.
This formulation balances link quality against inter-user spatial compatibility, promoting interference-aware scheduling decisions.
To maintain linearity, we introduce an auxiliary binary variable $v_{a,u,\up}$ to represent the product $x_{a,u} x_{a,\up}$, and enforce the equivalence using the linear constraints.
The sensing utility function is defined to capture the average strength of the bistatic links between illuminating and receiving APs for scheduled targets:
\begin{align}
    U_\text{sens} &= \sum_{t\in\setT} \lambda_t\, s_t \sum_{a_r\in\setA} \sum_{a_t\in\setA} y_{a_t, t}^\text{tx}\, y_{t,a_r}^\text{rx}\, \bar{g}_{a_t, t,a_r} \nonumber \\
    &\triangleq \sum_{t\in\setT} \sum_{a_r\in\setA} \sum_{a_t\in\setA} \lambda_t\, z_{a_t, t, a_r}\, \bar{g}_{a_t,t, a_r},
    \label{eq:usens}
\end{align}
where $z_{a_t, t, a_r} \triangleq s_t \, w_{a_t, t, a_r}$ and $w_{a_t, t, a_r} \triangleq y_{a_t, t}^\text{tx}\, y_{t,a_r}^\text{rx}$ are the auxiliary variables used for linearization. To ensure that~\eqref{eq:ucomm} and~\eqref{eq:usens} are comparable in magnitude, each is normalized by a reference value: $\bar{U}_\text{comm} = U_\text{comm} / U_\text{comm}^{(ref)}$ and $\bar{U}_\text{sens} = U_\text{sens} / U_\text{sens}^{(ref)}$, where $U_\text{comm}^{(ref)}$ and $U_\text{sens}^{(ref)}$ are computed under ideal conditions where all available resources are fully utilized. 

\subsection{MILP Formulation}

Based on the defined variables and utility functions, the association optimization problem is formulated as
\begin{subequations}\label{eq:opt_problem}
\begin{alignat}{2}
\hspace{-2mm} \max_{x, y, z, \tau, s, w, v}  & \alpha \bar{U}_\text{comm} + (1 - \alpha) \bar{U}_\text{sens} + \sum_{a\in\setA} \mu_a \tau_a \label{eq:opt_problem_obj}\\
\text{s.t.} \quad &\textstyle x, y, z, \tau, s, w, v \in \setBin \\
&\textstyle x_{a,u} \leq \tau_a &&\forall a, u \label{eq:opt_problem_c2} \\
&\textstyle y_{a,t}^{\text{tx}} \leq \tau_a &&\forall a, t \label{eq:opt_problem_c3-1} \\
&\textstyle y_{a,t}^{\text{rx}} \leq 1 - \tau_a &&\forall a, t \label{eq:opt_problem_c3-2} \\
&\textstyle \sum_u x_{a,u} + \sum_t y_{a,t}^{\text{tx}} \leq N_{\text{RF}}^{(a)} \quad && \forall a \label{eq:opt_problem_c4} \\
&\textstyle x_{a,u} + x_{a,\up} \leq 2 + \rho_\text{th} - \rhoauupbar \quad && \forall a, u, \up \label{eq:opt_problem_c5} \\
&\textstyle \sum_a y_{a,t}^{\text{tx}} \leq (|\setA| - 1) s_t && \forall t \label{eq:opt_problem_c6-1} \\
&\textstyle \sum_a y_{a,t}^{\text{rx}} \leq (|\setA| - 1) s_t && \forall t \label{eq:opt_problem_c6-2} \\ 
&\textstyle \sum_a y_{a,t}^{\text{tx}} \geq s_t && \forall t \label{eq:opt_problem_c7-1} \\
&\textstyle \sum_a y_{a,t}^{\text{rx}} \geq s_t && \forall t \label{eq:opt_problem_c7-2} \\
&\textstyle \sum_a y_{a,t}^\text{tx} \leq K_\text{tx} s_t && \forall t \label{eq:opt_problem_c8-1} \\
&\textstyle \sum_a y_{a,t}^\text{rx} \leq K_\text{rx} s_t && \forall t \label{eq:opt_problem_c8-2} \\
&\textstyle \sum_t y_{a,t}^\text{rx} \leq C_a^\text{rx} && \forall a \label{eq:opt_problem_c9} \\ \displaybreak \nonumber \\
\nonumber\\
& &&\hspace{-5.2cm} \raisebox{-3ex}[0pt][0pt]{\rotatebox{90}{\scriptsize \text{$x_{a,u} \cdot x_{a,\up}$}}} \left\{
\begin{alignedat}{2} \label{eq:opt_problem_c10}
&\textstyle v_{a,u,\up} \leq x_{a,u} \hspace{26mm} &&\forall a, u, \up \\
&\textstyle v_{a,u,\up} \leq x_{a,\up}  &&\forall a, u, \up \\
&\textstyle v_{a,u,\up} \geq x_{a,u} + x_{a,\up} - 1  \quad &&\forall a, u, \up \end{alignedat} \right. \\
& &&\hspace{-5.2cm} \raisebox{-4ex}[0pt][0pt]{\rotatebox{90}{\scriptsize \text{$s_t \cdot w_{a_t, t, a_r}$}}} \left\{
\begin{alignedat}{2} \label{eq:opt_problem_c11}
&\textstyle z_{a_t, t, a_r} \leq s_t \hspace{28.6mm} && \forall a_t, t, a_r \\
&\textstyle z_{a_t, t, a_r} \leq w_{a_t, t, a_r} && \forall a_t, t, a_r \\
&\textstyle z_{a_t, t, a_r} \geq s_t + w_{a_t, t, a_r} - 1 && \forall a_t, t, a_r \end{alignedat} \right. \\
& &&\hspace{-5.3cm} \raisebox{-4ex}[0pt][0pt]{\rotatebox{90}{\scriptsize \text{$y_{a_t, t}^\text{tx} \cdot y_{t, a_r}^\text{rx}$}}} \left\{
\begin{alignedat}{2} \label{eq:opt_problem_c12}
&\textstyle w_{a_t, t, a_r} \leq y_{a_t,t}^{\text{tx}} \hspace{23.85mm} && \forall a_t, t, a_r \\
&\textstyle w_{a_t, t, a_r} \leq y_{t,a_r}^{\text{rx}} && \forall a_t, t, a_r \\
&\textstyle w_{a_t, t, a_r} \geq y_{a_t,t}^{\text{tx}} + y_{t,a_r}^{\text{rx}}  - 1 && \forall a_t, t, a_r. \end{alignedat} \right.
\end{alignat}
\end{subequations}
The objective function~\eqref{eq:opt_problem_obj} is formulated as a weighted sum of the normalized communication and sensing utilities, controlled by a trade-off parameter $\alpha$.
A mode selection reward term is also incorporated, defined as $\sum_a \mu_a \tau_a$, where $\mu_a \ge 0$ is a tunable weight assigned to AP $a$.
This term biases the optimization toward activating a preferred subset of transmitting APs and enables flexible control over network behavior.
By tuning $\mu_a$, the system can encourage energy-efficient scheduling, limit unnecessary interference, and support high-level policies such as load balancing or hardware duty cycling.
In this work, $\mu_a$ is computed as a communication-readiness score, $\mu_a = \sum_u \lambda_u (\bar{g}_{a,u} / \bar{g}^\text{max})$, which quantifies the relative cost of switching AP $a$ to receive mode, normalized by the maximum observed channel gain $\bar{g}^\text{max}$.

In problem~\eqref{eq:opt_problem}, constraints~\eqref{eq:opt_problem_c2} and~\eqref{eq:opt_problem_c3-1} ensure that APs can serve CUs and illuminate targets only in transmit mode, and~\eqref{eq:opt_problem_c3-2} ensures only APs in receive mode can receive target echos.
Constraint~\eqref{eq:opt_problem_c4} enforces that the total number of entities (CUs and targets) served by AP $a$ does not exceed its number of RF chains.
Constraint~\eqref{eq:opt_problem_c5} introduces a correlation threshold $\rho_\text{th}$, and enforces that for any pair of CUs, if their channel correlation exceeds $\rho_\text{th}$, then at most one of them is allowed to be scheduled to avoid excessive interference.
Constraints~\eqref{eq:opt_problem_c6-1} and~\eqref{eq:opt_problem_c6-2} guarantee that if a target $t$ is scheduled, it is associated to at most $\Nap - 1$ transmit (receive) APs, thereby reserving at least one AP to operate in receive (transmit) mode, and constraints~\eqref{eq:opt_problem_c7-1} and~\eqref{eq:opt_problem_c7-2} further ensure that for any scheduled target, at least one AP is assigned as a transmitter and receiver, respectively.
Constraints~\eqref{eq:opt_problem_c8-1} and~\eqref{eq:opt_problem_c8-2} provide an optional cap on the number of APs associated with a scheduled target as transmitter or receiver, respectively.
Similarly, constraint~\eqref{eq:opt_problem_c9} also optionally limits the number of targets that a receiver AP $a$ can listen to by introducing a maximum capacity $C_a^\text{rx}$.
Constraints~\eqref{eq:opt_problem_c10}, \eqref{eq:opt_problem_c11}, and \eqref{eq:opt_problem_c12} are introduced to linearize the bilinear products of binary decision variables. Although this transformation increases the problem size by adding extra variables and constraints, it preserves the linear structure of the optimization problem, which is advantageous for enabling globally optimal solutions.
Since the objective~\eqref{eq:opt_problem_obj} and constraints~\eqref{eq:opt_problem_c2}–\eqref{eq:opt_problem_c12} are all linear, the optimization problem~\eqref{eq:opt_problem} is a mixed-integer linear program (MILP) and can be efficiently solved using standard MILP solvers. If the problem is feasible, the obtained solution is guaranteed to be globally optimal.


\section{Learning-Based ASSociation and ENTity Selection: ASSENT}
\label{sec:learning}


Solving the MILP introduced in~\eqref{eq:opt_problem} yields globally optimal association decisions, but at the expense of high computational complexity, especially in large-scale networks. 
In practice, real-time decision-making is critical, as user mobility, target dynamics, and time-varying channels demand rapid adaptation.
Furthermore, under fronthaul capacity constraints, it is impractical for the CS to frequently solve a large MILP using updated network-wide information.
Building upon the MILP, this section introduces ASSENT (learning-based ASSociation and ENTity selection), a GNN-based framework that learns the optimal association and mode-selection decisions derived from the MILP.
Rather than repeatedly solving the optimization problem at each scheduling interval, ASSENT employs supervised learning to map the lightweight statistics reported by APs to near-optimal decisions. 
Once trained, the model enables real-time and scalable inference with negligible computational cost at runtime, making it well-suited for distributed cell-free ISAC deployments under constrained fronthaul capacity.
The following subsections detail the dataset generation and GNN architecture used to implement the proposed framework.

\subsection{Dataset Generation}
\label{subsec:dataset}

To train the proposed ASSENT model, we generate a comprehensive dataset of association and mode-selection decisions by solving the MILP in~\eqref{eq:opt_problem} over a large number of independent channel realizations.
For each realization, the input system parameters summarized in Table~\ref{tab:dataset_params} include the communication channel gains $\mathbf{G}_\text{comm}$, inter-user spatial correlations $\mathbf{S}_\text{comm}$, and the bistatic sensing link gains $\mathbf{G}_\text{sens}$, as well as the control parameters $\alpha$, $\boldsymbol{\lambda}_\text{cu}$, and $\boldsymbol{\lambda}_\text{tg}$.
The MILP is solved centrally at the CS to obtain the globally optimal binary decision variables $\{\tau, \, x, \, s, \, y^\text{tx}, \, y^\text{rx}\}$.
Each realization therefore provides a complete input-output pair used for supervised learning.

\renewcommand{\arraystretch}{1.25}
\begin{table}[t]
\centering
\caption{Input and Output Parameters in ASSENT Dataset}
\label{tab:dataset_params}
\fontsize{8}{9}\selectfont
\begin{tabular}{|c|c|c|}
\hline
\textbf{Symbol} & \textbf{Definition} & \textbf{Size / Range} \\ \hline \hline
$\mathbf{G}_\text{comm}$ & Comm. channel gains (AP-CU) & $\mathbb{R}^{\Nap \times \Ncu}$ \\ \hline
$\mathbf{S}_\text{comm}$ & CU-CU spatial correlations & $\mathbb{R}^{\Nap \times \Ncu \times \Ncu}$ \\ \hline
$\mathbf{G}_\text{sens}$ & Sensing gains (Tx-Rx-Target) & $\mathbb{R}^{\Nap \times \Nap \times \Ntg}$ \\ \hline
$\alpha$ & Comm-Sensing trade-off & $[0,1]$ \\ \hline
$\boldsymbol{\lambda}_\text{cu}$ & Vector of CU priority weights & $\mathbb{R}^{\Ncu}$ \\ \hline
$\boldsymbol{\lambda}_\text{tg}$ & Vector of target priority weights & $\mathbb{R}^{\Ntg}$ \\ \hline \hline

$\boldsymbol{\tau}$ & AP mode (1 = Tx, 0 = Rx) & $\{0,1\}^{\Nap}$ \\ \hline
$\mathbf{X}$ & AP-CU association & $\{0,1\}^{\Nap \times \Ncu}$ \\ \hline
$\mathbf{s}$ & Target scheduling & $\{0,1\}^{\Ntg}$ \\ \hline
$\mathbf{Y}^\text{tx}$ & TxAP-Target association & $\{0,1\}^{\Nap \times \Ntg}$ \\ \hline
$\mathbf{Y}^\text{rx}$ & RxAP-Target association & $\{0,1\}^{\Nap \times \Ntg}$ \\ \hline
\end{tabular}

\vspace{-2mm}
\end{table}

To generate channel realizations, we consider a network with $\Nap=8$ APs, $\Ncu=10$ CUs, and $\Ntg=4$ targets.
The APs are uniformly placed on a circle of radius $350$ m, while users and targets are randomly positioned within a $1$ km $\times$ $1$ km area centered around the APs.
Each AP is equipped with $N_\text{RF}=4$ RF chains and a UCA with $M=16$ antennas. The carrier frequency is $3$ GHz, and the trade-off parameter $\alpha$ is randomly drawn from $[0,1]$.
Targets are modeled as cars with RCS statistics drawn from~\cite{nextGalliance_tr_phase2} and velocities uniformly distributed in $[0,10]$ m/s. The target RCS is modeled as a class-dependent random variable with mean value $\bar{\beta}_\text{class} \in [10, 20]$ dBsm.
To capture fluctuations due to the aspect angle between the target and the APs, a random variable $\Delta_\text{asp} \sim \mathcal{N}(0,\,\sigma_{\text{asp}}^2(v))$ is introduced, whose STD $\sigma_{\text{asp}}(v)$ varies with the target velocity $v$ as
\begin{equation}
    \sigma_\text{asp}(v) = \sigma_\text{asp}^\text{min} + (\sigma_\text{asp}^\text{max} - \sigma_\text{asp}^\text{min}) \frac{\log_{10}(1+v/v_\text{ref})}{\log_{10}(1+v_\text{max}/v_\text{ref})},
\end{equation}
where $v_\text{ref}$ is the reference velocity sensitivity (set to $1\,\text{m/s}$) and $v_\text{max}$ is the maximum velocity for the target class (set to $30\,\text{m/s}$).
The tunable parameters $\sigma_\text{asp}^\text{min}$ and $\sigma_\text{asp}^\text{max}$ control the impact of velocity on RCS fluctuations. Accordingly, each realization of the RCS for the AP-target link is expressed as
$\beta_{a,t}^{\text{rcs}}\,[\text{dB}] = \bar{\beta}_\text{class} + \Delta_\text{asp}$.
The priority weights $\lambda_u$ and $\lambda_t$ for each CU $u$ and target $t$ are uniformly sampled from $[\lambda_{\min}, \lambda^{\max}]$ for $\lambda_{\min} = 0.5$ and $\lambda^{\max}=\{5, 10\}$.
Using these parameters, a total of 10,000 independent realizations are generated by randomizing user and target positions, propagation conditions, and control parameters.
The dataset is divided into training, validation, and test subsets with an 80, 10, 10 ratio.
Each realization is serialized as a heterogeneous graph object and stored in binary format for efficient loading during training.
This dataset captures a statistically representative set of scenarios, enabling the learning model to generalize to unseen network configurations and trade-off conditions.

\subsection{GNN Architecture}

Modeling the ISAC network as a graph provides a natural way to represent its heterogeneous structure, where APs, CUs, and targets act as distinct node types connected through communication and sensing edges.
This formulation enables GNNs to capture both local spatial relationships and global coordination among entities, allowing the model to infer association and mode-selection patterns that generalize across network realizations.
Each node type employs a lightweight multilayer perceptron (MLP) encoder to map its input features to a shared latent space of dimension $H$.
Specifically, AP, user, and target features are embedded by MLPs with ReLU activation and dropout, followed by Layer Normalization to stabilize training and balance feature scales.
The encoded representations are then propagated through stacked HeteroConv layers that perform relation-specific message passing between node types.
Three message-passing operators were investigated: (i) NNConv, which uses learned edge-conditioned kernels, (ii) GATv2Conv, which introduces attention-based weighting for improved feature selection, and (iii) TransformerConv, which further models multi-head contextual dependencies.
These variants allow evaluation of the trade-off between computational efficiency and the model capacity to learn intricate interaction patterns within the ASSENT framework.

For prediction, dedicated MLP-based heads are used for each task: AP-CU associations ($\mathbf{X}$) on communication edges, AP mode selection ($\boldsymbol{\tau}$) on AP nodes, target scheduling ($\mathbf{s}$) on target nodes, and AP-target links ($\mathbf{Y}^\text{tx}$, $\mathbf{Y}^\text{rx}$) on sensing edges.
To jointly optimize all tasks, learnable log-variance parameters ($\log\sigma^2$) are introduced to adaptively weight each loss component, following the principle of task-uncertainty balancing.
The overall loss combines binary cross-entropy terms for each output with additional regularization penalties enforcing AP RF chain limits and coupling consistency between variables (e.g., $\tau_a$ and $y_{a,t}^\text{rx}$).
This formulation ensures stable multi-task learning and maintains consistency with the physical constraints of the original MILP problem.
Once trained, the proposed ASSENT model efficiently infers near-optimal association and mode-selection decisions with sub-millisecond latency, offering a scalable and practical solution for real-time operation in distributed cell-free ISAC systems.


\section{Evaluation}
\label{sec:evaluation}

\begin{figure}[!t]
    \centering
    \includegraphics[width=3.493in]{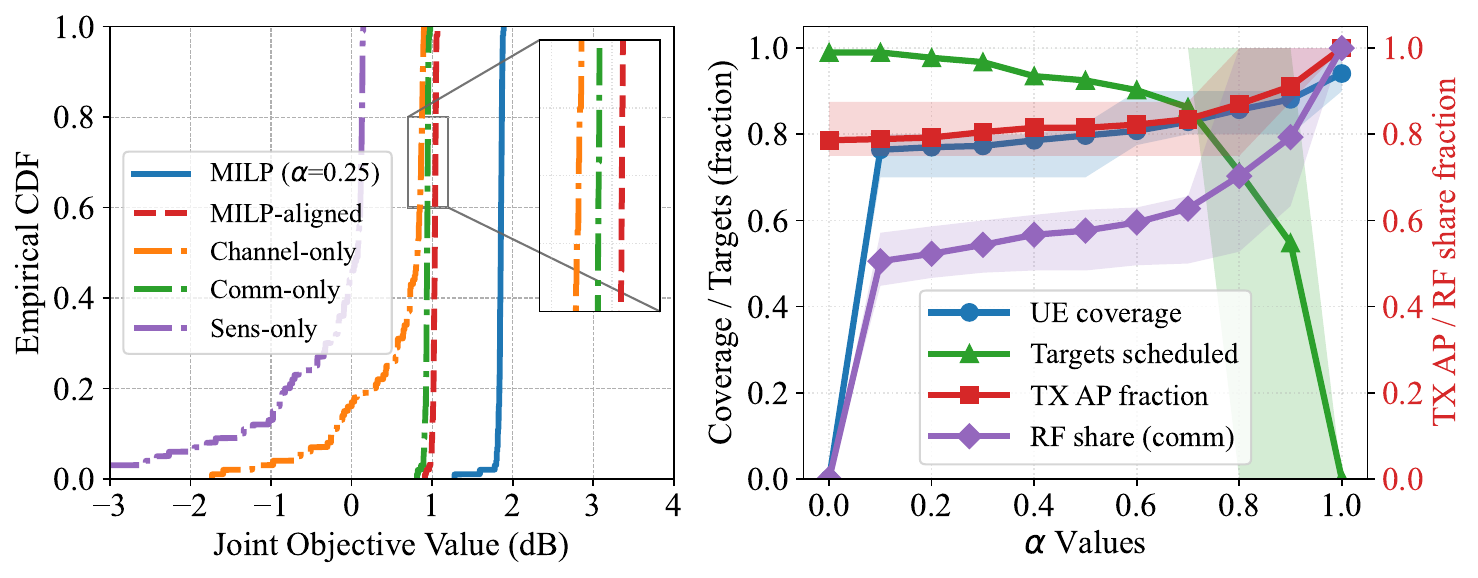}
    \caption{Performance of the MILP-based optimization~\eqref{eq:opt_problem}. (Left) Empirical CDF of the MILP utility and four greedy baselines: MILP-aligned (schedules exactly the same number of users and targets as MILP but greedily), channel-only (greedy only on channel gain), comm-only (scheduling only CUs greedily), and sens-only (scheduling only targets). (Right) Variation of network metrics: UE coverage fraction, scheduled-target fraction, Tx-AP fraction, and communication RF-share versus the trade-off parameter $\alpha$. Shaded band denotes the inter-quartile range (25th-75th percentile).}
    \label{fig:milp}
    \vspace{-2mm}
\end{figure}

We first evaluate the MILP-based association optimization against four greedy baselines: \textit{MILP-aligned, channel-only, communication-only, and sensing-only}.
Here, “greedy” denotes myopic selection based on the maximum instantaneous link strength (product of priority weight and channel gain for each entity).
For this evaluation, the priority weights $ \lambda_u$ and $\lambda_t$ are set to one, and for each value of the trade-off parameter $\alpha \in [0,1]$, a total of $10,000$ independent channel realizations are generated.
Fig.~\ref{fig:milp} (left) presents the empirical CDF of the achieved utilities for the MILP and greedy solutions, while Fig.~\ref{fig:milp} (right) illustrates how key network-level indicators, including fractions of UE coverage, scheduled-targets, transmit APs, and communication RF chain share, evolve as $\alpha$ varies.
These results confirm that the MILP formulation achieves the best utility and more balanced communication-sensing behavior than heuristic alternatives for various $\alpha$. 

\begin{figure}[!t]
    \centering
    \includegraphics[width=3.4935in]{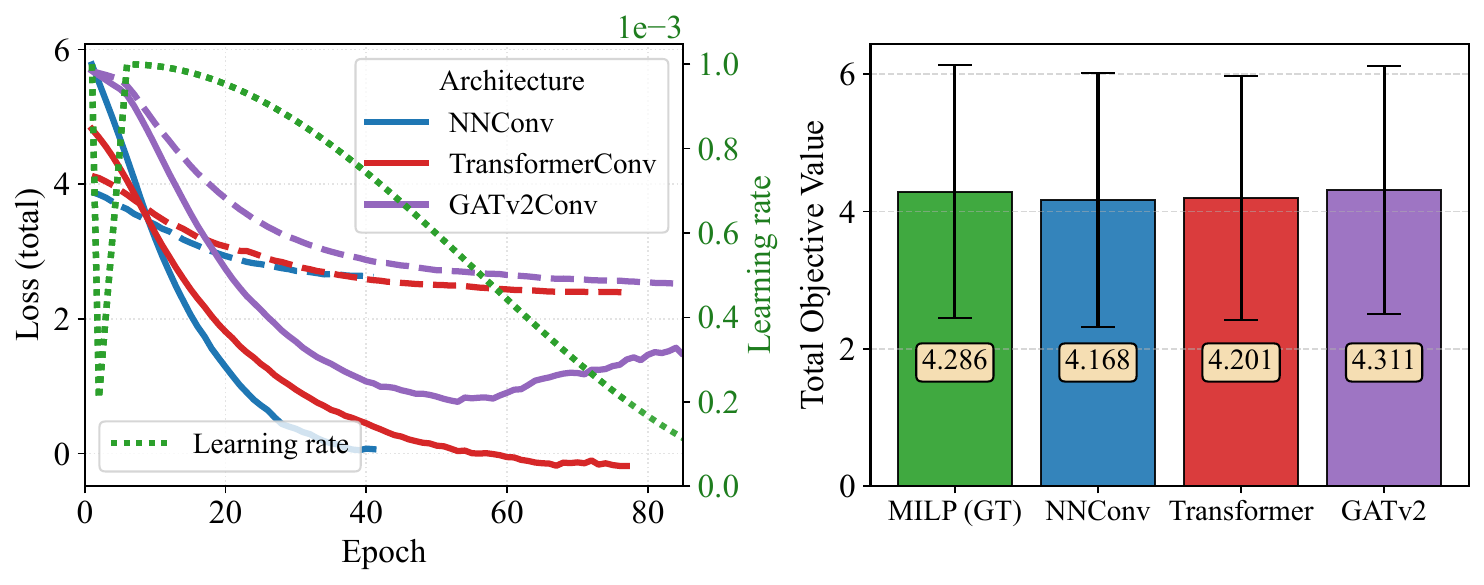}
    \caption{Performance of ASSENT. (Left) Training (solid) and validation (dashed) loss for three GNN architectures, with learning-rate schedule shown on the right y-axis. (Right) Average objective values on the test set achieved by each GNN architecture compared with MILP benchmark.}
    \label{fig:learning}
\end{figure}

\renewcommand{\arraystretch}{1.15}
\begin{table}[t]
\vspace{-2mm}
\centering
\scriptsize
\caption{Performance metrics of learning models on test set.}
\label{tab:metrics}
\begin{tabular}{|c|c|ccccc|}
\hline
\textbf{Model} & \textbf{Metric} & $\boldsymbol{\tau}$ & $\mathbf{X}$ & $\mathbf{s}$ & $\mathbf{Y}^\text{tx}$ & $\mathbf{Y}^\text{rx}$ \\
\hline \hline
\multicolumn{2}{|c|}{\textbf{N (samples)}} & 8000 & 80000 & 4000 & 32000 & 32000 \\
\hline
\multirow{8}{*}{NNConv}
& F1        & \textbf{0.985} & 0.907 & 0.975 & 0.841 & \textbf{0.919} \\
& Precision & 0.988 & 0.907 & \textbf{0.985} & 0.845 & \textbf{0.926} \\
& Recall    & \textbf{0.983} & 0.907 & 0.965 & 0.837 & 0.912 \\
& Brier     & \textbf{0.028} & 0.041 & 0.047 & 0.046 & \textbf{0.022} \\
\cline{2-7}
& TP        & 6637 & 17576 & 3070 & 5121 & 4502 \\
& FP        & 81  & 1800  & 47  & 940  & 360  \\
& TN        & 1164 & 58819 & 771 & 24941 & 26702 \\
& FN        & 118  & 1805  & 112  & 998  & 436  \\
\hline
\multirow{8}{*}{TransformerConv}
& F1        & 0.979 & \textbf{0.928} & 0.978 & \textbf{0.868} & 0.911 \\
& Precision & \textbf{0.994} & \textbf{0.948} & 0.974 & \textbf{0.873} & 0.871 \\
& Recall    & 0.964 & \textbf{0.909} & \textbf{0.982} & \textbf{0.862} & \textbf{0.956} \\
& Brier     & 0.045 & \textbf{0.029} & \textbf{0.030} & \textbf{0.038} & 0.027 \\
\cline{2-7}
& TP        & 6518 & 17650 & 3089 & 5262 & 4693 \\
& FP        & 37  & 959  & 84  & 763  & 694  \\
& TN        & 1199 & 59623 & 769 & 25136 & 26395 \\
& FN        & 246  & 1768  & 58  & 839  & 218  \\
\hline
\multirow{8}{*}{GATv2Conv}
& F1        & 0.983 & 0.908 & \textbf{0.981} & 0.848 & 0.918 \\
& Precision & 0.993 & 0.916 & 0.982 & 0.853 & 0.898 \\
& Recall    & 0.972 & 0.901 & 0.979 & 0.843 & 0.939 \\
& Brier     & 0.036 & 0.038 & \textbf{0.030} & 0.044 & 0.023 \\
\cline{2-7}
& TP        & 6589 & 17634 & 3081 & 5111 & 4566 \\
& FP        & 45  & 1613  & 56  & 884  & 517  \\
& TN        & 1177 & 58813 & 797 & 25052 & 26621 \\
& FN        & 189  & 1940  & 66  & 953  & 296  \\
\hline
\end{tabular}
\vspace{-2mm}
\end{table}

Next we evaluate the ASSENT learning framework using the dataset generated from the MILP solutions.
Unlike the fixed-parameter MILP evaluation, here $\alpha$, $\boldsymbol{\lambda}_{\text{cu}}$, and $\boldsymbol{\lambda}_{\text{tg}}$ are input variables and randomly sampled in the dataset generation process as discussed in Section~\ref{subsec:dataset}.
The three GNN variants NNConv, TransformerConv, and GATv2Con are implemented with hidden dimension $H=128$, three message-passing layers, batch size $16$, initial learning rate $10^{-3}$, cosine learning-rate scheduler, Adam optimizer, and early stopping. 
Fig.~\ref{fig:learning} (left) shows the training and validation loss trends with the learning-rate profile, and Fig.~\ref{fig:learning} (right) compares achieved objective values on the test set against the MILP benchmark.
Table~\ref{tab:metrics} summarizes detailed test-set performance metrics, including F1-score, precision, recall, and Brier score, along with true positive/negative (TP/TN) and false positive/negative (FP/FN) values for all five output variables.
The high F1-scores (above $0.84$ for all variables) demonstrate that ASSENT can accurately reproduce binary association and scheduling patterns learned from the MILP labels.
Precision and recall remain well balanced, confirming that the model neither over-activates nor under-selects AP-entity links.

All codes, datasets, and trained checkpoints used to produce the above results are available in our open-source simulation repository~\cite{zafari2025assentrepo}.
Overall, the experiments show that: (i) MILP yields consistent gains over greedy heuristics while revealing how network-level indicators trade off with $\alpha$, and (ii) ASSENT learns from MILP-optimal labels to deliver near-optimal utility and accurate decision variables on unseen test scenarios.
Given that inference is a single forward pass, the learned models provide a practical surrogate for optimization, supporting real-time operation in distributed ISAC systems. 


\section{Conclusion}

In this paper we have presented ASSENT (learning-based ASSociation and ENTity selection), a GNN-based framework for efficient association optimization in distributed cell-free ISAC systems. We formulated the association problem as a MILP that captures interference, mode selection, and RF chain constraints, and used its optimal solutions to supervise a heterogeneous GNN that infers association decisions from lightweight link statistics.
Simulation results demonstrate that ASSENT achieves near-MILP utility while accurately reproducing the optimal association patterns.
Once trained, a single forward pass efficiently replaces repeated MILP optimizations for real-time decision-making.
All code and datasets have been released to support reproducibility and further research.
Future work includes reinforcement learning for continual adaptation and experimental validation on distributed ISAC testbeds.

\bibliographystyle{IEEEtran}
\bibliography{ref}

@ARTICLE{liu2022dual-functional,
    author={Liu, Fan and Cui, Yuanhao and Masouros, Christos and Xu, Jie and Han, Tony Xiao and Eldar, Yonina C. and Buzzi, Stefano},
    journal={IEEE Journal on Selected Areas in Communications}, 
    title={{Integrated Sensing and Communications: Toward Dual-Functional Wireless Networks for 6G and Beyond}}, 
    year={2022},
    volume={40},
    number={6},
    pages={1728-1767},
    doi={10.1109/JSAC.2022.3156632}
}

@ARTICLE{ngo2017cellfree,
    author={Ngo, Hien Quoc and Ashikhmin, Alexei and Yang, Hong and Larsson, Erik G. and Marzetta, Thomas L.},
    journal={IEEE Transactions on Wireless Communications}, 
    title={{Cell-Free Massive MIMO Versus Small Cells}}, 
    year={2017},
    volume={16},
    number={3},
    pages={1834-1850},
    doi={10.1109/TWC.2017.2655515}
}

@ARTICLE{umut2025cellfreeisac,
    author={Demirhan, Umut and Alkhateeb, Ahmed},
    journal={IEEE Transactions on Communications}, 
    title={{Cell-Free ISAC MIMO Systems: Joint Sensing and Communication Beamforming}}, 
    year={2025},
    volume={73},
    number={6},
    pages={4454-4468},
    doi={10.1109/TCOMM.2024.3490740}
}

@INPROCEEDINGS{zafari2024admm,
    author={Zafari, Mehdi and Pandey, Divyanshu and Doost-Mohammady, Rahman and Uribe, César A.},
    booktitle={Proc. 58th Asilomar Conf. on Sig., Sys., and Comp.}, 
    title={{ADMM for Downlink Beamforming in Cell-Free Massive MIMO Systems}}, 
    year={2024},
    volume={},
    number={},
    pages={623-628},
    doi={10.1109/IEEECONF60004.2024.10943106}
}

@misc{zafari2025decentralized,
    title={{Coordinated Decentralized Resource Optimization for Cell-Free ISAC Systems}}, 
    author={Mehdi Zafari and Rang Liu and A. Lee Swindlehurst},
    year={2025},
    eprint={2508.01044},
    archivePrefix={arXiv},
    primaryClass={eess.SP},
    url={https://arxiv.org/abs/2508.01044}
}

@techreport{3gpp_r18,
    author = {{3GPP}},
    title        = {{Study on channel model for frequencies from 0.5 to 100 GHz (Release 18)}},
    institution  = {3rd Generation Partnership Project (3GPP)},
    type         = {3GPP TR},
    number       = {38.901},
    version      = {V18.0.0},
    year         = {2024},
    month        = {May},
    note         = {{ETSI, Sophia Antipolis, France}}
}

@techreport{nextGalliance_tr_phase2,
  title        = {{Channel Measurements and Modeling for Joint /Integrated Communication and Sensing, as well as 7--24 GHz Communication Channels, Phase II}},
  author       = {{Next G Alliance}},
  institution  = {ATIS Next G Alliance},
  year         = {2025},
  month        = {August},
  note         = {{White Paper}}
}

@misc{peng2025graph,
      title={{Graph Learning for Cooperative Cell-Free ISAC Systems: From Optimization to Estimation}}, 
      author={Peng Jiang and Ming Li and Rang Liu and Qian Liu},
      year={2025},
      eprint={2507.06612},
      archivePrefix={arXiv},
      primaryClass={eess.SP},
      url={https://arxiv.org/abs/2507.06612}, 
}

@ARTICLE{time-domain-scheduling-2024,
  author={Memisoglu, Ebubekir and Janjua, Muhammad Bilal and Arslan, Hüseyin},
  journal={IEEE Wireless Communications Letters}, 
  title={{Power-Efficient Time-Domain Scheduling for ISAC Beamforming}}, 
  year={2024},
  volume={13},
  number={10},
  pages={2837-2841},
  doi={10.1109/LWC.2024.3448528}
}

@ARTICLE{setareh2025scheduling,
  author={Abanto-Leon, Luis F. and Maghsudi, Setareh},
  journal={IEEE Trans. Veh. Technol.}, 
  title={{Optimal User and Target Scheduling, User-Target Pairing, and Low-Resolution Phase-Only Beamforming for ISAC Systems}}, 
  year={2025},
  volume={74},
  number={6},
  doi={10.1109/TVT.2025.3532915}
}

@inproceedings{wang2024GNN,
    author = {Wang, Zihuan and Wong, Vincent},
    title = {{Heterogeneous Graph Neural Network for Cooperative ISAC Beamforming in Cell-Free MIMO Systems}},
    year = {2024},
    isbn = {9798400704895},
    publisher = {Association for Computing Machinery},
    doi = {10.1145/3636534.3698223},
    booktitle = {Proc. 30th Annual Int. Conf. on Mobile Computing and Networking (MobiCom)},
    pages = {2161–2172},
    numpages = {12},
    location = {Washington D.C., DC, USA}
}

@misc{zafari2025assentrepo,
    author = {Mehdi Zafari},
    title = {{ASSENT-CellFree-ISAC: Simulation Framework}},
    year = {2025},
    howpublished = {GitHub repository},
    url = {https://github.com/LS-Wireless/ASSENT-CellFree-ISAC},
    note = {Accessed: November 10, 2025},
    version = {1.0.0}
}

\end{document}